\documentstyle{mn}
\input{psfig.sty}
\begin{document}

\title{Evolution of active galactic nuclei broad-line region clouds: 
       low- and high-ionization lines}
\author[D.R. Gon\c calves, A.C.S. Fria\c ca and V. Jatenco-Pereira]
      {D.R. Gon\c calves$^{1,2}$, A.C.S. Fria\c ca$^{2}$ 
      and V. Jatenco-Pereira$^{2}$ \\
$^{1}$ Instituto de Astrof\'\i sica de Canarias, C. Via L\'actea s/n, 
      E-38200 La Laguna - Tenerife, Spain \\
$^{2}$ Instituto Astron\^omico e Geof\'\i sico - USP, Av. Miguel Stefano 4200,
04301-904 S\~ao Paulo - SP, Brazil}
\date{denise@ll.iac.es, amancio@iagusp.usp.br, jatenco@iagusp.usp.br}
\maketitle

\begin{abstract}
The formation of quasar broad-line region (BLR) clouds via thermal instability
in the presence of Alfv\'{e}n heating has been discussed by Gon\c {c}alves et
al.  (1993a, 1996).  In particular, these studies showed the relevance of
Alfv\'{e}n heating in establishing the stability of BLR clouds in the intercloud
medium.  The present paper shows the results of time-dependent calculations (we
use a time-dependent hydrodynamic code) following the evolution of BLR clouds,
since their formation from the $10^7$ K intercloud medium.  We also calculate
the UV and optical line emission associated with the clouds in order to compare
with observations.  Our results are compared with those of UV and optical
monitoring of well-studied AGN, which suggest that the BLR is most probably
composed of at least two different regions, each one giving rise to a kind of
line variability, since low- and high-ionization lines present different
patterns of variability.  
We discuss the alternative scenario in which lines of different ionization
could be formed at the same place but heated/excited by distinct mechanisms,
considering as a non-radiative mechanism the Alfv\'{e}n heating.
\end{abstract}


\section{Introduction}

The study of the broad-line regions (BLR) of active galactic nuclei (AGN) is
crucial in understanding the central engine of these objects.   
In addition, as the BLR reprocesses energy emitted
by the continuum source at ionizing ultraviolet energies that cannot be observed
directly, the monitoring of emission lines provides unique pieces of information
on the central continuum source.
Photoionization by continuum radiation from the central source is almost
certainly the most important heating/ionizing mechanism in BLRs.  
One should note
that non-thermal processes, for instance incoherent synchrotron radiation, were
earlier invoked to explain AGN spectra (for instance, Rees 1984, 1987; Contini
\& Viegas-Aldrovandi 1990).  
In any case, the
pure photoionization model has a number of inconsistencies 
--- the `energy budget problem', the `line ratio problem' 
and the `line variation problem' (Dumont, Collin-Souffrin \& Nazarova 1998).  
In fact, Dumont et al.  (1998) speculate on
the need and nature of a non-radiative heating source of energy --- dissipation
of sound or Alfv\'en waves, weak shocks, magnetic dissipation, etc., as a
possibility to partially solve these problems.  The non-radiative process
discussed here is the interaction between Alfv\'{e}n waves, which results in a
dissipation process heating the gas.  
The present study aims at evaluating the
effects of Alfvenic heating (AH) as non-radiative heating, in addition to pure
photoionization, on the broad-line spectrum.

We studied two damping mechanisms for Alfv\'en waves: the nonlinear and turbulent. 
These damping mechanisms have been investigated before, showing 
the relevance of Alfv\'en waves and their dissipation processes in all the 
following astrophysical objects. Late-type and proto stars 
(Jatenco-Pereira \& Opher 1989a, b; Vasconcelos, Jatenco-Pereira \& Opher 2000);
solar wind 
(Jatenco-Pereira \& Opher 1989c; Jatenco-Pereira, Opher \& Yamamoto 1994);
galactic and extragalactic jets 
(Opher \& Pereira 1986; Gon\c calves, Jatenco-Pereira \& Opher 1993b);
early-type stars (dos Santos, Jatenco-Pereira \& Opher 1993; Gon\c calves, 
Jatenco-Pereira \& Opher 1998); broad line regions of
quasars (Gon\c calves, Jatenco-Pereira \& Opher 1993a, 1996); and finally, 
optical filaments in cooling flows (Fria\c ca et al. 1997).

The scenario for BLRs considered in this article is the two-phase BLR, in which
the optical--UV lines are emitted by clouds embedded in a hot, external,
intercloud medium (Krolik, McKee \& Tarter 1981).  However, thermal pressure
alone is not enough to confine BLR clouds if the external medium is heated only
by Compton scattering (Mathews \& Ferland 1987; Mathews \& Doane 1990).  Mathews
\& Doane (1990) take into consideration various physical processes in the
heating--cooling the BLR, including:  i) recombination heating and cooling; ii)
radiative losses in the range between $10^4$~K to $3 \times 10^7$~K due to the
electron excitation of resonance transitions in the commonest metal ions; iii)
thermal Bremsstrahlung; and iv) Compton heating--cooling.  They conclude that
thermal instability cannot form the quasar line-emitting clouds for the observed
pressure $P \sim 10^{-2}$~dyne~cm$^{-2}$.  On the other hand, magnetic fields
and intense radiation may provide an important contribution to the formation and
confinement of the clouds (Rees 1987; Gon\c calves et al. 1993a, 1996).  
The additional heat source considered here, AH, together with more
conventional ones (see Mathews \& Ferland 1987; Krolik 1988; Mathews \& Doane
1990), maintain the broad-line emitting clouds and the intercloud medium in
pressure equilibrium (Gon\c {c}alves et al.  1993a, 1996), as required in the
two-phase BLR model (see Netzer 1990).  Within the framework of our model, the
correlation between continuum and line emission of the BLR is also affected by
AH as an additional heating/excitation source of the BLR, acting together with
photoionization by the central source.

The present study extends the stationary equilibrium calculations 
of Gon\c {c}alves
et al.  (1993a, 1996, hereafter Papers I and II, respectively), in which it was
found that AH could establish a stable two-phase medium in BLRs, by performing
time-dependent calculations for the evolution of line-emitting clouds from their
formation from a $10^7$~K intercloud gas to $\sim 10^4$~K clouds.  We compare
the results of our models with observational data coming from optical and UV AGN
monitoring campaigns.

Section~2 reviews the knowledge of magnetic fields in BLRs and the AH
mechanisms.  The time-dependent models for BLR clouds are discussed in
Section~3.  Model results and their comparison with the observations of AGN UV
and optical lines are given in Section~4.  An overall discussion as well our
main conclusions are stated in Section~5.

\section{Alfv\'{e}n heating in the BLR intercloud medium}

The classical observational signature of magnetic fields in AGN is the radio
emission resulting from synchrotron radiation of relativistic electrons
spiralling in a magnetic field.  Rees (1987), considering the magnetic
confinement of BLR clouds, has estimated a magnetic field strength of 
$\sim 1$~G, in the inner parsec of AGN (this value refers to the hot intercloud gas).
Since magnetic field are present in BLRs, we also expect the existence of
Alfv\'{e}n waves.  On disturbing a uniform magnetic field, the magnetic tension
makes a wave propagate along the field lines with the Alfv\'{e}n speed, 
$v_{\rm A}$ (for a review see Priest 1994).  The perturbations of the magnetic fields
could originate in the central region of the AGN, which is highly perturbed as
we infer from the fast continuum variability.  Therefore, two conditions that
are fulfilled by the central region of AGN make them favourable sites for the
generation of Alfv\'{e}n waves:  they contain magnetic fields and are highly
perturbed.

The diagnosis of the basic physical parameters of BLRs is not simple because
electron densities are high enough for almost all forbidden lines to be
collisionally suppressed.  However, the temperature is estimated via similarity
with other ionized gases, with temperatures of about $10^4$~K.  The
characteristic broad-line widths are $\sim 5000$~km~s$^{-1}$, but the above
temperature implies a line-of-sight velocity dispersion of $\sim
10$~km~s$^{-1}$!  Therefore, line widths do not reflect thermal motions only,
but are also due to bulk motions of individual broad-line emitting clouds.
These clouds are `usually assumed, without justification, to have no turbulent
velocity' (cf.  Baldwin 1997).  On the other hand, the clouds could not be
highly turbulent since in this case the effects on the line asymmetries would be
enormous.  The blue and red asymmetries of the broad lines are usually explained
by infall or outflow, but we are still far from a complete explanation of these
features (see Espey 1997 for a brief review of velocity diagnostics in BLRs).
Therefore, the line widths reveal the existence of bulk motions and, possibly,
of turbulence in the BLR, thus fulfilling the conditions for the BLR to produce
Alfv\'{e}n waves.

Once we admit the existence of Alfv\'{e}n waves in the AGN BLRs, 
the Alfv\'{e}n waves are subject to several damping processes.
In this work we consider the nonlinear and the turbulent damping processes.
Assuming spherical geometry for the clouds, we present below
the dependence of the Alfv\'en wave heating rate on density and temperature
(see Papers I and II for a detailed discussion). 

High amplitude waves cause nonlinear mode couplings and, 
in this process, the resulting modes are rapidly damped. 
In general, the waves that result from the damping processes are sound 
waves; thus, the nonlinear Alfvenic energy is converted into thermal energy. 
An Alfv\'{e}n wave is likely
to dissipate because of its nonlinear interaction with either a non-uniform
ambient field or another Alfv\'{e}n wave (Wentzel 1974).  The nonlinear
interaction of magnetohydrodynamic waves has been treated in detail by Kaburaki
\& Uchida (1971), Chin \& Wentzel (1972) and Uchida \& Kaburaki (1974). 
In regions of strong ($v_{\rm A}>c_{\rm s}$) magnetic field, 
one Alfv\'en wave can decay into another 
Alfv\'en wave and a sound wave travelling in the opposite direction. 
The resulting Alfv\'en 
wave has a frequency smaller than the original one and it can, 
in turn, decay 
into another lower--frequency Alfv\'en wave plus an acoustic wave. 
The cascade continues until all the Alfvenic energy has been 
converted to acoustic waves that dissipate rapidly.
Considering this mechanism, in our models, the onset of AH occurs for
$\beta=(B^2/8\pi )/nk_{\rm B}T > \beta_{\rm on} > 1$,
where $\beta_{\rm on}$ is the initial value of $\beta$ when AH becomes important
($\beta >1$ corresponds to $v_{\rm A}/c_{\rm s}>\sqrt{6/5}$ for an ideal 
$\gamma=5/3$ gas, where $c_{\rm s}$ is the sound speed). 
Note that our definition of $\beta$ corresponds, in a plane parallel geometry,
to $\beta = p_{\rm B}/p_{\rm gas}$, 
whereas in plasma physics, $\beta_{\rm pl} = p_{\rm gas}/p_{\rm B}$ denotes the 
``$\beta$ parameter" of the plasma.
The present convention for $\beta$ has been
widely used in cooling flow studies (e.g. David \& Bregman 1989),
and has been employed by Fria\c ca et al. (1997)
in the study of the effects of AH in optical filaments in
cooling flows.

According to Lagage \& Cesarsky (1983), the nonlinear damping rate is

\begin{equation}
\Gamma _{\rm nl}=\frac 14\sqrt{\frac \pi 2}\xi \bar{\omega}\left( \frac{c_{\rm s}}{v_{\rm A}}%
\right) \frac{\rho \langle \delta v^2\rangle }{B^2/8\pi }\,,
\end{equation}

\noindent 
where $\xi =5-10$, $\bar{\omega}$ is the characteristic Alfv\'{e}n frequency and
$\rho \langle \delta v^2\rangle /(B^2/8\pi )$ the ratio of the energy density of
Alfv\'{e}n waves to that of the magnetic field. 
In laboratory plasmas, the cut-off frequency of Alfv\'en waves is often
comparable to the ion-cyclotron frequency, $\omega_{\rm ci}=e B/m_{\rm i} c$,
so we can write $\bar{\omega}$ as a fraction $F$ of $\omega_{\rm ci}$,
i.e., $\bar{\omega}=F \omega_{\rm ci}$.
In addition, there are experimental grounds to set $F \approx 0.1$
or smaller (Burke, Maggs \& Morales 1998).

In terms of AH, e.g.  
$H_{\rm nl}$ (erg~cm$^{-3}$~s$^{-1}$), we have:  
$H_{\rm nl}={(\Phi_{\rm w}\,\Gamma_{\rm nl}})/{v_{\rm A}}$, where 
$\Phi _{\rm w}$ is the wave flux, $\Gamma _{\rm nl}$
is given by (1), and $v_{\rm A}$ is $B/{\sqrt{4\pi \rho }}$.  As the
perturbation collapses, the component of the magnetic field perpendicular to the
direction of compression is amplified.  For spherical geometry $B\propto
A^{-1}\propto \rho ^{2/3}$, where $A$ is the cross-sectional area perpendicular
to the magnetic field.  As a consequence, $v_{\rm A}\propto \rho ^{1/6}$.  Let
the dependence of $\Phi _{\rm w}$ on $\rho $ be given by 
$\Phi _{\rm w}\propto A^{-1}\propto \rho ^{2/3}$.  Taking 
$\Phi _{\rm w}=\rho \langle \delta v^2\rangle v_{\rm A}$, we have 
$\rho \langle \delta v^2\rangle \propto \rho^{1/2}$ .  Since 
$c_{\rm s}\propto T^{1/2}$, for $\bar{w}$ independent of the
density, we obtain:

\begin{equation}
H_{\rm nl}\propto \rho ^{-1/2}\,T^{1/2}\,.
\end{equation}

In addition, the nonlinear interaction between outward and inward propagating
waves results in an energy cascading process (Tu, Pu \& Wei 1984).  Following
Hollweg (1986, 1987), the volumetric heating rate associated with the cascade
process (e.g.  associated with turbulent damping of Alfv\'{e}n waves) can be
written as

\begin{equation}
H_{\rm tu}=\rho {\langle \delta v^2\rangle }^{3/2}/L_{{\rm corr}}\,\,,
\end{equation}

\noindent
where $L_{{\rm corr}}$ is a measure of the transverse correlation length of the
magnetic field.  Again, for spherical geometry, 
$\langle \delta v^2\rangle \propto \rho ^{1/2-1}\propto \rho ^{-1/2}$, and 
adopting $L_{{\rm corr}}\propto B^{-1/2}\propto \rho ^{-1/3}$, we obtain:

\begin{equation}
H_{\rm tu}\propto \rho ^{7/12}\,.
\end{equation}

We would like to call the reader's attention to the fact that Paper II has a
misprint in equations (4) and (6), more specifically, in the dependence of 
$\Phi_{\rm w}$ on the density (in the calculations the correct expression was used).
Equations (2) and (4) now exhibit the right dependence of $\Phi _{\rm w}$ on the
density.  As a matter of fact, these heuristically derived density dependencies
were included in Paper II, since we have discused not only AH density dependency
due to nonlinear, resonance surface and turbulent Alfvenic dampings, but also a
range of other density dependencies for ``like-Alfvenic heatings'' (see Fig.  9
of Paper II).  In the other two papers discussing AH in other environments than
those of BLRs, the correct expressions for $H_{\rm nl}$ and $H_{\rm tu}$ are 
found (Fria\c {c}a et al.  1997; Gon\c {c}alves, Jatenco-Pereira \& Opher 1998).

\section{Time-dependent cooling condensations}

We have investigated the evolution of quasar BLR clouds from their formation out
of the hot phase of the BLR and calculated their ultraviolet and optical
signatures within the scenario outlined in Section~2.  The evolution of the
cooling clouds is obtained by solving the hydrodynamic equations for mass,
momentum and energy conservation (see Fria\c {c}a 1986, 1993):

\begin{equation}
\frac{\partial \rho }{\partial t}+\frac 1{r^2}\frac \partial {\partial
r}\left( r^2\rho u\right) = 0
\end{equation}

\begin{equation}
\frac{\partial u}{\partial t}+u\frac{\partial u}{\partial r}=-\frac{1}{\rho} 
\frac{\partial p_{\rm t}}{\partial r}-\frac{GM(r)}{r^2}
\end{equation}

\begin{eqnarray}
\frac{\partial U}{\partial t}+u\frac{\partial U}{\partial r}
&=&\frac{p_{\rm t}}{\rho
^2} \left( \frac{\partial \rho}{\partial t}+u\frac{\partial \rho }{\partial r}
\right) -\Lambda \rho + \frac{H(\rho,T)}{\rho} \nonumber \\ 
& &\mbox{} +\frac{1}{\rho r^2} \frac{\partial}{\partial r}
\left(\kappa r^2 \frac{\partial T}{\partial r} \right)
\end{eqnarray}

\noindent where $u$, $\rho $, $p_{\rm t}$, $U$, $T$ and $\kappa$ are the gas
velocity, density, total pressure, the specific internal energy, 
temperature, and the thermal
conductivity coefficient.  The equation of state ,

\begin{equation}
U=\frac 32\frac{k_{\rm B}T}{\mu m_{\rm H}} +\frac{u_{\rm B}}{\rho}
\end{equation}

\noindent 
relates $U$ to $T$ and $u_{\rm B}=B^2/8\pi$,
the internal magnetic energy density
($k_{\rm B}$ is the Boltzmann's constant, $m_{\rm H}$ is the hydrogen atomic
mass and $\mu =0.62$ is the mean molecular weight of a fully ionized gas with
solar abundances).

The total pressure, $p_{\rm t}$, is the sum of the thermal and magnetic
pressures, $p_{\rm t}=p+p_{\rm B}$.  
The constraints on the magnetic pressure come
from considerations on the $\beta$-parameter, assuming that the magnetic
pressure of the unperturbed medium would not exceed its thermal pressure, in
accordance with Mathews \& Doane (1990).  We assume frozen-in fields tangled on
scales less than that of the perturbation, so that, for spherical collapse, 
$B \propto \rho^{2/3}$ ($p_{B} \propto \rho^{4/3}$).
In this case the magnetic field can be considered as effectively isotropic
and we treat the isotropic tangled magnetic field
as a $\gamma=4/3$ gas, and so $p_{\rm B}=u_{\rm B}/3=B^2/24\pi$.

The self-gravity of the condensations is taken into account
through $M(r)$, the mass distribution of the gas.
Note, however, that the self-gravity is
much less important than the tidal force due to the BH,
as one can see from the comparison between
the surface gravity of the perturbation
(of mass $M=4\pi \rho_{\rm m} R^3$ and radius $R=L/2$
for the values of $\rho_{\rm m}$ and $L$ defined in the next section)
$$g_{\rm suf}=G M/R^2= 8.4 \times 10^{-8} \; {\rm cm} \, {\rm s}^{-1}$$
\noindent
and the tidal acceleration due to a central supermassive
black hole of mass $M_{\rm BH}=10^8$ M$_{\odot}$
at a distance $r=30$ light-days of the perturbation:
\noindent
$$g_{t\rm id}=G M_{\rm BH} R/r^3= 0.71 \; {\rm cm} \, {\rm s}^{-1}.$$

In the energy equation, the heating term $H(\rho,T)$
follows either equation (2)
or equation (4) and it is turned off in the case of no heating.
In the adopted
definition of the cooling function $\Lambda(T)$, $\Lambda(T)\rho^2$, is the
cooling rate per unit volume.  Since there is no ionization equilibrium for
temperatures lower than $10^6$~K, the ionization state of the gas at $T<10^6$~K
is obtained by solving the time-dependent ionization equations for all ionic
species of H, He, C, N, O, Ne, Mg, Si, S, Ar and Fe.  The radiative cooling
function was obtained via the collisional ionization code MEKA (Mewe,
Gronenschild \& van der Oord 1985).  The coefficients of collisional ionization,
recombination and charge exchange of the ionization equations is also from MEKA.
The adopted abundances are cosmic abundances usually assumed in AGN models
(Netzer 1990).  If solar abundances are used, they are taken from Grevesse \&
Anders (1989).

The last term of eq.  7 accounts for the thermal conductivity.  We take the
thermal conductivity coefficient, $\kappa$, as a fraction, $\eta$, of the
Spitzer (1956) value $\kappa_{\rm S}=4.87 \times 10^{-7}$~erg~cm$^{-1}$~K$^{-1}$, with
$\eta=10^{-4}$.  The thermal conductivity should be inhibited, otherwise it
would prevent the development of the thermal instability.  The thermal
conductivity would erase any steep temperature gradient arising in the medium,
thus heating the cooler gas of BLRs to the higher temperatures of the
surrounding hot gas.  Therefore, the presence of colder clouds embedded in a hot
gas requires some reduction of the thermal conduction.  A similar inhibition of
thermal conduction seems to be needed also in cooling flows in clusters of
galaxies:  the classic (Spitzer) thermal conductivity would drive a heat flow
towards the cooling flow, and, as a consequence, there should be no cooling
flow!  The magnetic fields could be at the origin of the inhibition of thermal
conductivity.  In the presence of magnetic fields, the heat conduction is
suppressed in the direction perpendicular to the magnetic field, taking place
substantially only along magnetic field lines.  If the magnetic field is very
tangled, the heat conduction can be reduced by a large factor.  Other processes
of inhibition of thermal conductivity besides magnetic field entanglement are
possible.  For instance, in cooling flows in galactic clusters, electromagnetic
instabilities driven by temperature gradients (or electric currents in other
situations) can also cause this inhibition, even for non-tangled field lines
(Pistinner, Levinson \& Eichler 1996).

The spherically symmetric hydrodynamical equations are solved using a
finite-difference, implicit code based on Cloutman (1980).  The code is run in
the Lagrangian mode and the grid points are initially spaced logarithmically,
with a grid of 100 cells.  The artificial viscosity for the treatment of the
shocks follows the formulation of Tsharnutter \& Winkler (1979) based on the
Navier-Stokes equation.  In contrast with the von Neumann--Richtmyer artificial
viscosity (Richtmyer \& Morton 1967), this form of artificial viscosity vanishes
for homologous contraction.  Thermal conduction is included by
operator-splitting using a Crank--Nicholson scheme.  The outer boundary
conditions on pressure and density are derived by including an outer fictitious
cell, the density and pressure in which are obtained from extrapolation of power
laws over the radius fitted to the outermost real cells.

The initial density perturbations are characterized by an amplitude, $A$, and a
length scale, $L$, following

\begin{equation}
\frac{\delta \rho}{\rho} = \frac{A \sin(x)}{x} \,\,\,;\,\,\,\,\,\,\, 
x=\frac{2 \pi r}{L} \,\,\, .
\end{equation}

\noindent 
We have also assumed that the perturbations are isobaric and nonlinear 
($A$ =1), avoiding the uncertainties concerning the linear development of thermally
unstable modes in the presence of a gravitational field 
(David, Bregman \& Seab 1988).  
The unperturbed hot-phase intercloud medium is characterized by
a total hydrogen density $n_{\rm m}=5\times 10^6$~cm$^{-3}$ and 
a temperature $T_{\rm m}=10^7$~K, appropriate for the inner regions of
an AGN, $\sim 10^{18}$~cm from the central source, where lies the BLR. 
If the gas cools isobarically, when it reaches $T=10^4$ K,
the resulting central electronic density would be $6\times 10^9$ ~cm$^{-3}$,
consistent with the absence of forbidden line emission in AGN BLRs.
We fixed $\beta_0 = B^2/(8\pi n k_{\rm B} T_{\rm m}) = 0.03$ in accordance with Mathews \& Doane (1990)
who pointed out that this quantity should be between 0.003 and 0.03, in a similar calculation with no Alfvenic heating.
The
length scale of the perturbation was fixed at $5\times 10^{16}$~cm.  The
corresponding column density is $N_{\rm H} = 3.2 \times 10^{23}$~cm$^{-2}$.

Here we investigate several representative models with magnetic fields, one with
no AH and the others with AH due to two damping mechanisms (nonlinear and
turbulent) and at different efficiencies (see Table 1).  Model I allows for the
presence of a magnetic field only through a magnetic pressure term
in the equation of motion.  In models II and III, the energy
equation includes AH as an additional heating term.  We considered nonlinear and
turbulent Alfv\'{e}n wave heating, from the above equations (eqs.  (2) and (4)),
and these heating terms have the forms

\begin{equation}
H_{\rm nl}=\zeta \,H_0(n/n_0)^{-1/2}(T/T_0)^{1/2}\,\,,
\end{equation}

\noindent
for nonlinear heating, and 

\begin{equation}
H_{\rm tu}=\zeta \,H_0(n/n_0)^{7/12}\,\,,
\end{equation}

\noindent
for turbulent heating, where $\zeta$ is the efficiency of Alfv\'{e}n heating.
The choice of the normalization is the following:  1) $T_0=10^5$~K (the
temperature at which optical emission begins to be important); 2)
$n_0=13.59\times (5\times 10^6)$~cm$^{-3}$, where $13.59$ would be 
the compression factor from the unperturbed 
$n_{\rm m}=$ $5\times 10^6$~cm$^{-3}$ and 
$T_{\rm m}=10^7$~K initial state to $T=T_0$,
under isobaric conditions for a plane parallel geometry; and 3)
$H_0=(3/2)\,n_0k_{\rm B}T/t_{\rm col}$~erg~cm$^{-3}$~s$^{-1}$, where $t_{\rm col}$ is
the collapse time (defined below) for the cloud with no Alfvenic heating.

\section{Modelling results and data comparison}

\begin{table*} 
{\Large
\begin{minipage}{60mm}
\caption{Properties of the Models}
\begin{tabular}{|l|c|c|c|c|c}
\hline
{ Model} & { AH} & $\zeta $ & { t}$_{{\rm t,\ col}}$ \\
& & & $10^7{\rm s}$ \\ \hline
I & none & - & $9.88$ \\ 
IInl & nonlinear & $10\%$ & $10.10$ \\ 
IItu &  turbulent & $10\%$ & $9.89$ \\ 
IIInl & nonlinear & $30\%$ & $10.60$ \\ 
IIItu & turbulent & $30\%$ & $9.93$ \\ 
\hline
\end{tabular}
\end{minipage}
}
\end{table*}

We investigate models with magnetic field and AH (excepted model I in which
there is no AH) using $H_{\rm nl}$ and $H_{\rm tu}$ with the efficiency of the
conversion of Alfv\'{e}n wave energy in Alfvenic heating, $\zeta$, equivalent to
$10\%$ and $30\%$.  In model I the magnetic effect is considered only as
magnetic pressure in the equation of motion.  All the other
models contain AH as an additional heating term in the energy equation.  The
characteristics of our models are summarized in Table~1:  model identification
(column 1); the Alfvenic heating considered in each model (column 2); the
efficiency $\zeta$ of the AH (column 3); and the total collapse time 
$t_{\rm {t,col}}$, i.e.  the time since the beginning of the calculations until 
the time when the innermost cell reaches $8\times 10^3$~K (column 4).

First of all, note that AH can maintain stable equilibrium between the optical
line-emitting cloud and the intercloud medium.  In fact, Fria\c ca et al.
(1997), studying the Alfv\'en heating/excitation of the optical filaments in
cooling flows, have found that one of their models (with nonlinear heating and
$\zeta = 100\%$) reached thermal stability, showing that AH can suppress strong
growth of thermal instabilities in cooling flows.  This suggests that AH could
induce a stable two-phase equilibrium in other astrophysical environments, such
as the BLRs of AGN, which were investigated with simple time-independent
calculations in Papers I and II.  Therefore, we take the initial conditions of
our simulations from Paper II.  As we see from Table 1, an increase in the AH
efficiency from $10\%$ to $30\%$, leads to longer total collapse times:  from
10.10 to 10.60 $\times 10^7$~s for the models with $H_{\rm nl}$ and from 9.89 to
9.93 $\times 10^7$~s for those with $H_{\rm tu}$.  The same trend of collapse
time with $\zeta$ was found by Fria\c ca et al.  (1997), thus confirming that AH
inhibits development of thermal instabilities.

\begin{figure}
\begin{center}
\vbox{
\psfig{file=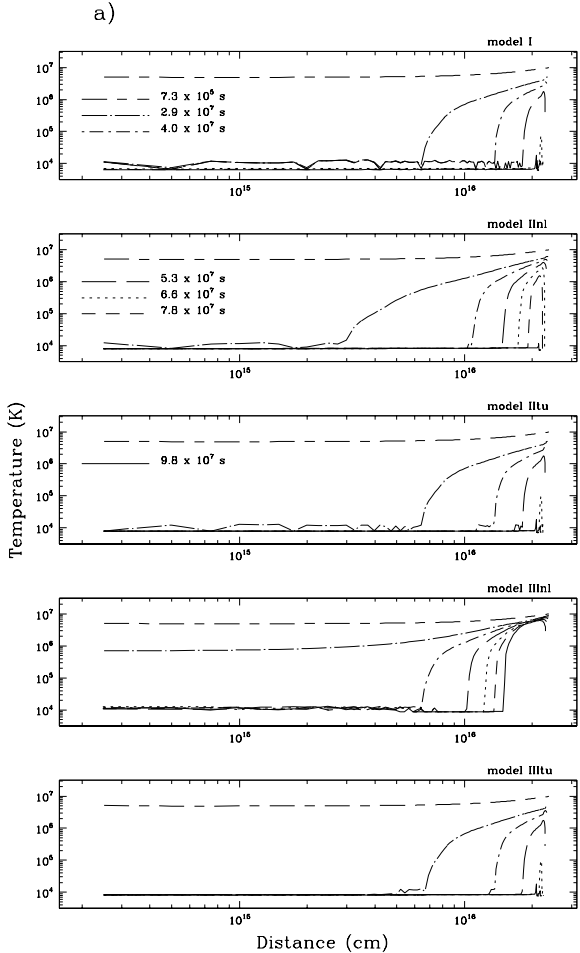,width=8.0truecm,bbllx=133pt,bblly=166pt,bburx=470pt,bbury=695pt}}
\end{center}
\caption[]{Profiles of {\bf (a)} temperature for models I(no AH), 
IInl($H_{\rm nl}$, $\zeta=0.1$), IItu($H_{\rm tu}$, $\zeta=0.1$), IIInl($H_{\rm nl}$, 
$\zeta=0.3$) and IIItu($H_{\rm tu}$, $\zeta=0.3$) for several evolution times.}
\end{figure}

Figure~1 -3 show the temperature, thermal pressure and $\beta$ (magnetic to
thermal pressure ratio) profiles of our models.  These profiles are shown from
the beginning of the calculations, throughout the optical phase 
($T\leq 10^5$~K), until $9.8 \times 10^7$~s (very close to the total collapse time of
model I, $9.88 \times 10^7$~s).  The influence of the AH on the collapse of the
perturbed gas can be well understood following the profiles of Figure 1.  For
model I (with no AH), nearly all the gas contained in the perturbation reaches
$\sim 10^4$~K in the total collapse time of $9.88 \times 10^7$~s.  All the
models with AH have total collapse times longer than this.  Comparing the
results for nonlinear and turbulent AH models, we can see that nonlinear AH is
more effective in inhibiting the development of the collapse than turbulent AH.
Note that model IIInl (nonlinear AH, with $\zeta=30\%$) has the highest total
collapse time, and therefore the smallest amount of ``cool'' gas when the
evolution time reaches $9.8 \times 10^7$~s.

As we see from the thermal pressure profiles of Figure~2, as compared to 
Figure~1, the drop in the temperature causes a reduction in the thermal pressure.
This behaviour is a consequence of the fact that the sound crossing time
is much longer than the cooling time, so there was no time for repressurizing.
The $\beta$ profiles (Figure~3) show that, at
the beginning of the calculations, the magnetic pressure is much smaller than
the thermal one; but, as the perturbation evolves, the magnetic pressure becomes
more important, as a result of the drop in the temperature of the perturbed gas,
$\beta$ becomes $>$ 1,
and any (slight) compression of the gas 
will contribute to the increase in $\beta$
(assuming frozen-in fields, the magnetic field increases 
according to $B \propto \rho^{2/3}$ for a spherical geometry).

When the temperature falls to $\sim 5\times 10^5$~K, the gas begins to emit
optical and UV lines.  Before this optical line emitting stage, the gas cooling
occurs via EUV and X-rays lines.  We have calculated the emissivity of
low-ionization optical lines as well as that of high-ionization UV lines.  The
results of our models may be compared to the observational data from monitoring
campaigns, such as those of the Seyfert 1 galaxies Fairall 9 (with {\it IUE},
Rodr\'\i guez-Pascual et al.  1997; in the optical range, Santos-Lle\'o et al.
1997) and NGC 5548 (Dumont et al.  1998).

We set the beginning of the phase of optical--UV line emission when the
temperature of the innermost cell (that is, the first cell to cool) falls below
$5\times 10^5$~K.  Figure~4 shows the evolution during this phase of the
luminosity of the low-ionization line H$\beta$, and of the high-ionization lines
He{\sc ii}~1640~\AA, C{\sc iv}~1549~\AA, N{\sc v}~1240~\AA \- and 
O{\sc vi}~1034~\AA, for the models I, IInl and IIInl.  The variability pattern of the
three models is very similar for all the lines studied.  In addition, AH
enhances significantly, albeit not strongly, the broad line emission.  We should
note that the amplitude of variation of the H$\beta$ line is smaller than that
of the high-ionization ions, and that the time-scale of its variability is much
longer.

\begin{figure}
\begin{center}
\vbox{
\psfig{file=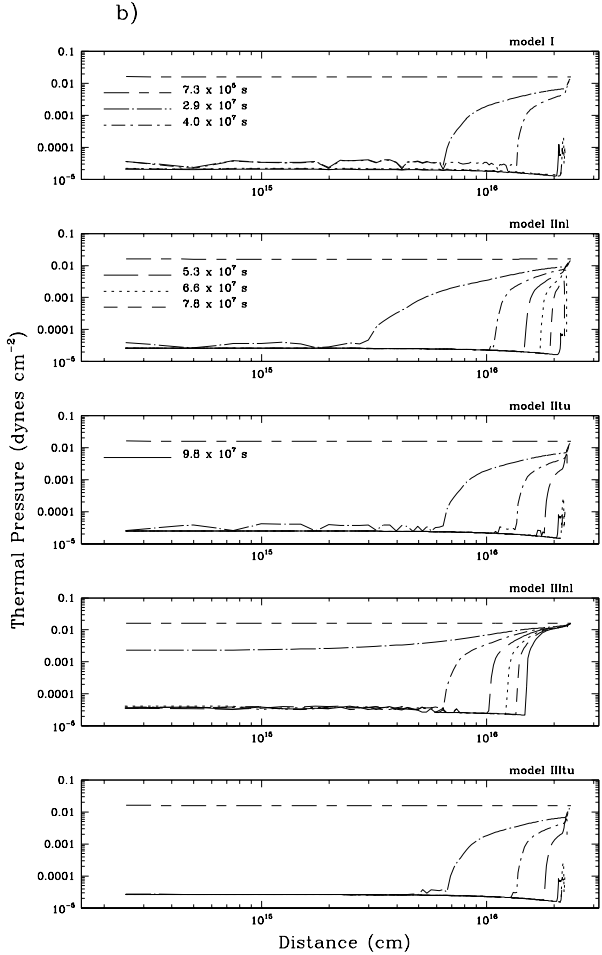,width=8.0truecm,bbllx=133pt,bblly=166pt,bburx=470pt,bbury=695pt}}
\end{center}
\caption[]{Profiles of {\bf (b)} thermal pressure for models I(no AH), 
IInl($H_{\rm nl}$, $\zeta=0.1$), IItu($H_{\rm tu}$, $\zeta=0.1$), IIInl($H_{\rm nl}$, 
$\zeta=0.3$) and IIItu($H_{\rm tu}$, $\zeta=0.3$) for several evolution times.}
\end{figure}

\begin{figure}
\begin{center}
\vbox{
\psfig{file=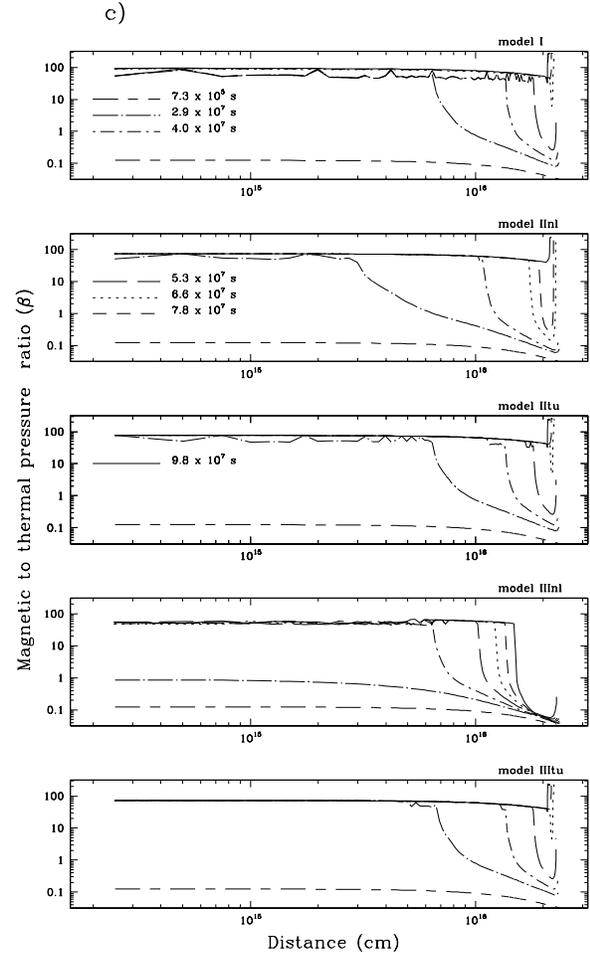,width=8.0truecm,bbllx=133pt,bblly=166pt,bburx=470pt,bbury=695pt}}
\end{center}
\caption[]{Profiles of {\bf (c)} magnetic to thermal pressures ratio for models I(no AH), 
IInl($H_{\rm nl}$, $\zeta=0.1$), IItu($H_{\rm tu}$, $\zeta=0.1$), IIInl($H_{\rm nl}$, 
$\zeta=0.3$) and IIItu($H_{\rm tu}$, $\zeta=0.3$) for several evolution times.}
\end{figure}

With respect to the variability of the collisionally excited high-ionization UV
lines, their luminosities show well-defined maxima occurring $\sim$ 10--25 days
after the appearance of the first gas with $T< 5\times 10^5$~K.  The
variability of the lines C{\sc iv}~1549~\AA, N{\sc v}~1240~\AA\ and 
O{\sc vi}~1034~\AA\ is faster and of larger amplitude than that of the 
He{\sc ii}~1640~\AA\ line.  
For model IIInl, the time interval $\Delta t_{1/2}$ for the line luminosity
to decrease from its maximum value to half-maximum value is
46.0~d, 24.2~d, and 31.9~d, for the lines C{\sc iv}~1549~\AA, 
N{\sc v}~1240~\AA\ and O{\sc vi}~1034~\AA, respectively, 
whereas for He{\sc ii}~1640~\AA,
$\Delta t_{1/2}=152$~d.  
Note that, in addition to increase the luminosities of the lines
(during the period from the time of peak luminosity and $t=100$ d, the increase
is $30-55\%$ for He{\sc ii}~1640~\AA,
$60-70\%$ for C{\sc iv}~1549~\AA,
$48-52\%$ for N{\sc v}~1240~\AA,
and $51-60\%$ for O{\sc vi}~1034~\AA)
the AH makes the luminosity peaks more prominent.

The evolution of the H$\beta$ luminosity contrasts with the evolution of the
high-ionization UV lines in that it is almost constant within an elapsed time of
about 300 days.  It shows a broad maximum at 150--200 days and varies within
only $10\%$ between 70 and 320 days.  The fact that the amplitude of variation
of the H$\beta$ line is smaller than that of the high-ionization ions, and that
the time-scale of its variability is much longer is in a first order agreement
with observations.  For Fairall 9, the H$\beta$ line shows a smoother pattern of
variability than He{\sc ii}~1640~\AA, with a time-lag $\tau_{\rm d}=23.0$~d and a
ratio of maximum to minimum fluxes $R_{\rm max}=1.55$, whereas for 
He{\sc ii}~1640~\AA, $\tau_{\rm d}=4.2$ d and $R_{\rm max}=3.25$.  However, a closer
comparison of the luminosity and time-lags between Balmer and high-ionization
lines reveals that the predicted luminosity of H$\beta$ is too low and its
variability too slow.  For model IIInl, at $t=150$~d, even close to the maximum
of the H$\beta$ line, and far from the maximum of the C{\sc iv}~1549~\AA\ line,
the ratio C{\sc iv}~1549~\AA/H$\beta$ is 24.7, whereas the typical value for AGN
BLRs is $\sim 5$.  
AH does not seem to account for the Balmer recombination
lines, which could be explained by photoionization, but it makes a significant
contribution, both to the energetics and to the variability of the
high-ionization lines.

\begin{figure}
\begin{center}
\vbox{
\psfig{file=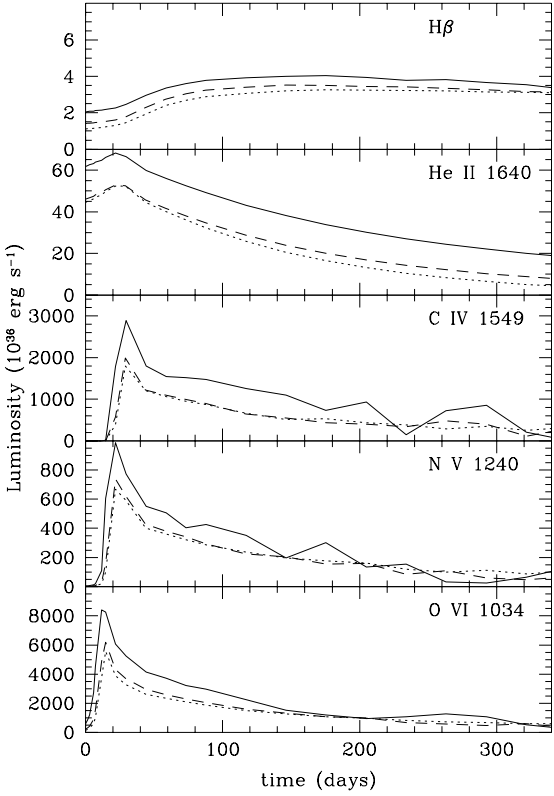,width=8.0truecm,bbllx=65pt,bblly=158pt,bburx=462pt,bbury=696pt}}
\end{center}
\caption[]{Luminosity variability of the line H$\beta$, as well as those 
of the UV high-ionization lines He{\sc ii}~1640~\AA,
C{\sc iv}~1549~\AA, N{\sc v}~1240~\AA\ and O{\sc vi}~1034~\AA, for 
models I (dotted lines), IInl (dashed lines) and IIInl (solid lines).
The time is counted from the beginning of the optical phase
(i.e. when $T$ becomes $<5\times 10^5$ K).}
\end{figure}

\section{Discussion and Conclusions}

In our model,
the values of the time of the luminosity peak $t_{\rm max}$
--- $t_{\rm max}=29.7$~d for C{\sc iv}~1549~\AA,
$t_{\rm max}=21.9$~d for N{\sc v}~1240~\AA,
and $t_{\rm max}=21.9$~d for He{\sc ii}~1640~\AA, for model IIInl --- 
reproduce the observed sequence of
time-lags in AGN BLRs, with the time lags for N{\sc v}~1240~\AA\ and 
He{\sc ii}~1640~\AA\ being of comparable magnitude and shorter than those of 
C{\sc iv}~1549~\AA\ (our models predicts an even shorter variability
time scale for O{\sc vi}~1034~\AA: $t_{\rm max}=11.7$~d for model IIInl).
The usual interpretation is that the He{\sc ii}--N{\sc v}
emission region is located at a shorter distance from the central continuum
source than the C{\sc iv}~1549~\AA\ region --- in the case of NGC 5548, at 6 and
10 light-days, respectively --- (Dumont et al.  1998).  
The present results
suggest that at least part of this trend could be due to different evolutionary
time scales for each line within one single cloud or cloud ensemble at one given
position with respect to the central source, instead of different clouds giving
rise predominantly to different lines at differing locations.

We can interpret our results in the light of the discussions by
Dumont et al. (1998), which, although give support to scenario
in which the BLR consists at least of two different regions,
in which low- and high-ionization lines are predominantly produced,  
also point out some inconsistencies of the photoionization models
based in this scenario with respect to the energy budget, the line ratios
and the line variability.
The variability represents the greater challenge for
the stratified model for the formation of the high- and low-ionization lines, 
in which the more ionized lines are formed closer to the center
(CIV 1549$\rm \AA$ and He II 1640$\rm \AA$) and the less ionized further out
(Collin--Souffrin et al.  1986; Collin--Souffrin, Hameury \& Joly 1988).

It is possible that the standard stratified model 
suffers of an excessive number of emission regions. 
For NGC 5548, Dumont et al. (1998) distinguishes four
zones: the MgII-CIII] emission zone, at $\sim 30-50$ lt-days; 
the Balmer lines-HeI emission zone, at $\sim 12-20$ lt-days; 
the Ly$\alpha$-CIV emission zone, at $8-12$ lt-days; 
the NV-HeII emission zone, at $4$ lt-days.
Our results also allow for several locations for the
formation of the broad lines, with the high-ionization lines
being produced closer to the central continuum source,
and having an important contribution from AH,
and the Balmer lines being produced at a more distant location, and powered
mostly by photoionization, which would be responsible
for most of its variability.
Of course, some fraction of the broad-line
spectrum could be non-variable, and the AH-powered clouds
could contribute to a slowly-variable part of the 
Balmer lines emission.
The AH mechanism would be one more ingredient in
the constitution of the time-lags.  
The decreasing values of $t_{\rm max}$ 
for the lines C{\sc iv}~1549~\AA, 
N{\sc v}~1240~\AA\ and O{\sc vi}~1034~\AA\ shows that
different time lags do not necessarily indicate different line formation
regions.
A stratified model simpler than the standard one
could consider the formation of the low-ionization lines 
in clouds far from the center
accommodate the production of the highest ionization
lines (from C{\sc iv}~1549~\AA\ to O{\sc vi}~1034~\AA)
in the same location very close to the center
(e.g. in the case of NGC 5548, in the few inner lt-day), 
with part of delay of the lines with respect to fluctuations of
the continuum arising from evolutionary effects
within the clouds, and  heated/excited by
non-radiative processes, in our case, by AH.

The use of Alfv\'en heating in our models is not aimed at
solving all the problems pointed out by Dumont et al (1998):
the `energy budget problem', the `line ratio problem' 
and the `line variation problem'.
For example, AH is no solution to the `energy budget problem', 
because of the small contribution of
AH to H$\alpha$ emission (and to the Balmer series).
However, AH alleviates the `line variation problem'
because it reproduces the trend of time lags 
with different ionization lines,
with no need for an excessive stratification of clouds
emitting different lines.
Although AH does not account for the Balmer recombination
lines, which could be explained by photoionization,
it makes a significant contribution both to the energetics
and to the variability of the high-ionization lines. 
What we explored here is
the possibility that several species of high-ionization lines, are 
produced at the same location, however, 
being not only photoionized but also heated/excited by, in the present 
case, Alfv\'enic heating.

The lag between continuum and line variability could arise
from some process associated with maxima in the continuum emission 
initiating the collapse.
A possible mechanism for triggering the collapse of the perturbation
is Compton cooling.
Compton cooling resulting from luminous QSOs in clusters of galaxies
has been shown to be able to induce or sustaining cooling flows
(Fabian \& Crawford 1990; Atuko et al. 1998).
Compton cooling can also be important
in accretion disks (Esin 1997).
Mathews and Ferland (1987) find that the Compton temperatures
appropriate to quasar spectra are in the range
$2.8\times 10^6 < T_{\rm C} < 4.6\times 10^7$ K.
If the source radiation is soft 
($T_{\rm C} < T_{\rm m}$, the temperature of the intercloud medium),
burst in the central continuum could induce cooling on
time scales close to the Compton cooling time:
$$
t_{\rm C}=\frac{\frac 32 n k_{\rm B} T_{\rm m}}{4 \frac{k_{\rm B} \sigma_{\rm T}}{m_{\rm e} c^2}
\frac{L_{\rm bol}}{4\pi R^2} n (T_{\rm m}-T_{\rm C})}
=\frac 32 \pi \frac{m_e c^2}{\sigma_{\rm T}}\frac{R^2}{L_{\rm bol}}
(1-\frac{T_{\rm C}}{T_{\rm m}})^{-1},
$$
\par\noindent
where $R$ is the distance to the source and $L_{{\rm bol}}$ is its
bolometric luminosity 
(and the other symbols have their usual meaning).
For typical values of $R$ and $L_{{\rm bol}}$ in AGN BLRs,
$$
t_{\rm C}=4.06 \frac{(R/30 \;{\rm light\;days})^2}
{(L_{\rm bol}/10^{47}\;{\rm erg\,s}^{-1})} (1-\frac{T_{\rm C}}{T_{\rm m}})^{-1}
\; {\rm days},
$$
\par\noindent
so it is feasible to induce collapse of perturbations
on the required short time scales.

As mentioned in section 2, high amplitude waves cause nonlinear couplings and
the nonlinear Alfvenic energy is, in general, converted into thermal energy. 
The nonlinear damping rate (eq. (1)) is a function of $\bar{\omega}$, 
which can be written in terms of the ion-cyclotron frequency, 
$\bar{\omega} = F \omega_{\rm ci}$. 
We assume $F = 0.1$, which is consistent with laboratory experiments on
heat transport in magnetized plasmas. 
In a electron beam experiment (Burke et al. 1998), 
classical conduction is observed for the first 2 ms
since beam turn-on, after
which fluctuations with frequencies $\sim 0.1 \omega_{\rm ci}$ spontaneously grow,
and at 3 ms, large amplitude fluctuations 
with frequencies $ 0.02 \omega_{\rm ci}$ appear.
Following equation (1), the damping length $L_{\rm damp}=v_{\rm A}/\Gamma$
can be written as
$$
L_{\rm damp}=1.92\times 10^7 (\frac{\xi}{5})^{-1} 
(\frac{\rho \langle \delta v^2\rangle}{B^2/8\pi})^{-1}
F^{-1}n_{\rm H}^{-1/2}\beta^{1/2}\; {\rm cm}.$$
\noindent
For the onset of the AH ($\beta=1$, $n_{\rm H}=5\times 10^6$ cm$^{-3}$),
assuming $\xi=5$, $\rho \langle \delta v^2\rangle/(B^2/8\pi)=1$,
and $F=0.1$,
$L_{\rm damp} = 8.7 \times 10^4$ cm, much smaller compared with the initial dimension of the 
system ($5 \times 10^{16}$ cm) and the radius where the Alfvenic heating is acting 
($4.2 \times 10^{15}$ cm for the model IIInl). This implies that the waves are damped very 
rapidly and so must be generated locally in the cloud. We have evidence for generation in situ
of Alfv\'en waves from observations of outward and inward fluctuations propagation in the solar
wind (Coleman 1968; Belcher \& Davis 1971). Roberts et al. (1987) using data from
Helios and Voyager spacecraft, investigated the origin and evolution of low-frequency planetary
fluctuations from $0.3$ to $20 AU$. They concluded that the outward traveling fluctuations are
predominantly generated by the Sun, but that in situ turbulence, most likely due to stream shear,
generates fluctuations with both inward and outward senses of correlation. 
This mixed state 
provides stronger evidence for in situ generation. 
We assume that the above scenario can be applied to the BLR clouds 
allowing the in situ generation of Alfv\'en waves.

It would be interesting to compare the results of the present
model with other models assuming local heating sources.
Besides photoionization model by central continuum, local 
emission models have also been proposed in order to explain the 
$H{\alpha}$ luminosities and the major line ratios of cluster nebulae 
(Voit et al. 1994). 
The major difference between the work of 
Voit et al. (1994) and the present one is that we are dealing with an 
evolutionary model where the heating source is non-radiative (AH heating) 
and the total pressure includes both thermal and magnetic pressures. 
On the 
other hand, in the model of Voit et al. (1994) it is assumed a pressure 
and heating-cooling equilibrium, while our model is time-dependent and
does not assume any equilibrium.
Moreover, Voit et al. (1994) present the results of a photoionization
code (CLOUDY), considering the soft X-ray/EUV radiation field of
a cooling flow in which cold clouds are embedded,
while we restricted ourselves to purely collisional emissivity calculations.

Our model is not aimed 
at reproducing the continuum emission of any particular AGN. 
Instead is a tool to investigate separately from photonization effects
the role of magnetic fields (magnetic pressure and Alfv\'en heating) 
as a mechanism contributing to the delay between the continuum variability 
and the line variability. 
A complete model for the BLRs would include 
photoionization, shocks and magnetic effects. The focus of the present 
work is on the third ingredient.

We should also note that even non-standard radiative heating mechanisms
deserve proper consideration.
If the hot intercloud medium is turbulent,
mixing layers are expected to develop around the colder BLR clouds.
In the conditions of cooling flows,
Begelman \& Fabian (1990) have shown that EUV radiation from mixing layers
could power the line emission of optical filaments in cooling flows.
It is possible that, also in BLRs, the radiation from mixing layers
around the BLR clouds could constitute an additional, local,
photoionization source, besides the central continuum.

Other issue is the stability of the BLR cold clouds.
Several external forces are expected to be acting on (and against) 
the BLR clouds in the harsh environment constituted by the intercloud
medium --- high temperatures ($T\sim 10^{7-8}$ K),
high relative velocities (up to $\approx 10^4$ km s$^{-1}$),
and, probably, turbulence.
Significant motions of the BLR clouds relative to the intercloud medium
could lead to disruption of the clouds, and radiation forces could
ablate the protoclouds before they contribute fully to the
broad line emission (Mathews \& Doane 1990). 
Also with respect to this topic,
magnetic fields have a role to play,
providing supporting pressure and confinement of the clouds 
(Rees 1987; Gon\c calves et al. 1993, 1996).
In addition, in the motion of the BLR clouds through the whole of BLR
(with the accompanying effects on the line profiles)
magnetic fields could have a dynamical role, as
it seems to the case in cooling flows (Gon\c calves \& Fria\c ca 1999).
Finally, in order that the density reaches values
high enough to account for the absence of forbidden lines 
in AGN BLRs, some process should remove the magnetic field
in the late evolution of the perturbations,
when pressure equilibrium has been achieved,
otherwise the magnetic field would prevent further
compression of the BLR cloud.
One mechanism that was shown to be effective in optical filaments
in cooling flows is magnetic reconnection (MR)
(Jafelice \& Fria\c ca 1996, Fria\c ca \& Jafelice 1999).
Therefore, it would be interesting to investigate
MR in BLR clouds, both as a heating source 
and as mechanism for removing magnetic support.

\subsection*{Acknowledgements}

We are grateful to R. Mewe for making us available the MEKA code
and its atomic database. We thank the anonymous referee for his/her comments,
which helped us to significantly improve the paper. DRG would like
to thank the grant of the Brazilian agency FAPESP (98/7502-0) and the 
Spanish DGES PB97-1435-C02-01 grant. ACSF and VJP
would like to thank the Brazilian agency CNPq for partial support.  We also
acknowledge partial support by Pronex/FINEP (41.96.0908.00).

\vfill\eject


\begin{thebibliography}{99}

\bibitem{} Atuko N., Habe A., Isibasi N., 1998, MNRAS, 295, 632

\bibitem{b01}  Baldwin J.A., 1997, in Emission Lines in Active Galaxies: New
Methods and Techniques, IAU Col. 159, eds. B.M. Peterson, F.-Z. Cheng and
A.S. Wilson, ASP Conference Series, 113, p. 80

\bibitem{}  Begelman M.C., Fabian A.C., 1990, MNRAS, 244, 26

\bibitem{} Belcher J. W., Davis, L., 1971, J. Geophys. Res., 76, 3534

\bibitem{} Burke A. T., Maggs J. E., Morales G. J., 1998, PRL, 81, 3659

\bibitem{}  Chin Y., Wentzel D. G., 1972, Ap\&SS, 16, 465

\bibitem{} Cloutman L. D., 1980, Los Alamos Report. LA-8452-MS

\bibitem{} Coleman P. J., 1968, ApJ, 153, 371

\bibitem{}  Collin-Souffrin S., Joly M., Pequignot D., Dumont S., 1986, 
A\&A, 166, 27

\bibitem{}  Collin-Souffrin S., Hameury J. M., Joly M., 1988, A\&A, 205, 19

\bibitem{}  Contini M., Viegas-Aldrovandi S., 1990, ApJ, 350, 125

\bibitem{}  David L. P., \& Bregman J. N., 1989, ApJ, 337, 97

\bibitem{}  David L. P., Bregman J. N., \& Seab C. G., 1988, ApJ, 329, 66

\bibitem{} dos Santos L. C., Jatenco-Pereira V., Opher R. 1993, ApJ, 410, 732

\bibitem{}  Dumont A. M., Collin-Souffrin S., Nazarova L., 1998, A\&A, 331, 11

\bibitem{}  Espey B., 1997, in Emission Lines in Active Galaxies: New
Methods and Techniques, IAU Col. 159, eds. B.M. Peterson, F.-Z. Cheng and
A.S. Wilson, ASP Conference Series, 113, p. 175

\bibitem{}  Esin A.A., 1997, ApJ, 482, 400

\bibitem{} Fabian A.C., Crawford C.S., 1990, MNRAS, 247, 439

\bibitem{}  Fria\c {c}a A. C. S., 1986, A\&A, 164, 6

\bibitem{}  Fria\c {c}a A. C. S., 1993, A\&A, 269, 145

\bibitem{}  Fria\c {c}a A. C. S., Jafelice L. C., 1999, MNRAS, 302, 491

\bibitem{}  Fria\c {c}a A. C. S., Gon\c {c}alves D. R., Jafelice L. C.,
Jatenco-Pereira V., Opher R., 1997, A\&A, 324, 449

\bibitem{}  Gon\c {c}alves D. R., Fria\c {c}a A. C. S., 1999, MNRAS, 309, 651

\bibitem{}  Gon\c {c}alves D. R., Jatenco-Pereira V., Opher R., 1993a, Paper
I, ApJ, 414, 57

\bibitem{}  Gon\c {c}alves D. R., Jatenco-Pereira V., Opher R., 1993b, A\&A, 279, 351

\bibitem{}  Gon\c {c}alves D. R., Jatenco-Pereira V., Opher R., 1996,
Paper II, ApJ, 463, 489

\bibitem{}  Gon\c {c}alves D. R., Jatenco-Pereira V., Opher R., 1998, ApJ,
501, 797

\bibitem{}  Grevesse N., Anders E., 1989, Cosmic Abundances of Matter. Am. 
Inst. Phys., ed. C.J. Waddington, New York, p. 183

\bibitem{}  Hollweg J. V., 1986, J. Geophys. Res., 91, 4111

\bibitem{}  Hollweg J. V., 1987, ESA, in Proc. 21st ESLAB Symp. on 
Small--Scales Plasma Processes in Solar Chromosphere Corona, Interplanetary 
Medium and Planetary Magnetospheres, p. 161

\bibitem{} Jafelice L. C.,  Fria\c {c}a A. C. S., 1996, MNRAS, 280, 438

\bibitem{} Jatenco-Pereira V., \& Opher R., 1989a, A\&A, 209, 327

\bibitem{} Jatenco-Pereira V., \& Opher R., 1989b, MNRAS, 236, 1

\bibitem{} Jatenco-Pereira V., \& Opher R., 1989c, ApJ, 344, 513

\bibitem{} Jatenco-Pereira V., Opher R., \& Yamamoto L. C., 1994, ApJ, 432, 409

\bibitem{}  Kaburaki O., Uchida Y., 1971, PASPJ, 23, 405

\bibitem{}  Krolik J. H., 1988, ApJ, 325, 148

\bibitem{}  Krolik J. H., McKee C. F., \& Tarter C. B., 1981, ApJ, 249, 422

\bibitem{}  Lagage P. O., Cesarsky C. J., 1983, A\&A, 125, 249

\bibitem{}  Mathews W. G., Doane J. S., 1990, ApJ, 352, 423

\bibitem{}  Mathews W. G., Ferland G. J., 1987, ApJ, 323, 456

\bibitem{}  Mewe R., Gronenschild E. H. B., van der Oord G. H. J., 1985, A\&AS,
62, 197

\bibitem{}  Netzer H., 1990, in Active Galactic Nuclei, ed. J.J.--L
Courvoisier and M. Mayor (Berlin: Springer--Verlag) 

\bibitem{} Opher R., \& Pereira V. J. S., 1986, Astrophys. Lett., 25, 107

\bibitem{}  Priest E. R., 1994, in Plasma Astrophysics, ed. J.G. Kirk, D.B.
Melrose and E.R. Priest (Berlin: Springer--Verlag), p. 38

\bibitem{}  Pistinner S., Levinson A., \& Eichler D., 1996, ApJ, 467, 162

\bibitem{}  Rees M. J., 1984, ARA\&A, 22, 471

\bibitem{}  Rees M. J., 1987, MNRAS, 228, 47

\bibitem{} Roberts D. A. et al., 1987, J. Geophys. Res., 92,12023

\bibitem{}  Rodr\'\i guez-Pascual P. M., et al., 1997, ApJS, 110, 9

\bibitem{}  Rychtmyer R. D., \& Morton K. W., 1967, in Difference Methods for 
            Initial Value Problems, Interscience, New York

\bibitem{}  Santos-Lle\'o M., et al., 1997, ApJS, 112, 271

\bibitem{} Spitzer L., 1956, in Physics of Fully Ionized Gases, 
Interscience, New York

\bibitem{} Tscharnuter W. M., Winkler K. H., 1979, Comp. Phys. Comm., 18, 171

\bibitem{}  Tu C. Y., Pu Z. Y., Wei F. S., 1984, J. Geophys. Res., 89, 9695

\bibitem{}  Uchida Y., Kaburaki O., 1974, Sol. Phys., 35, 451

\bibitem{} Vasconcelos M. J., Jatenco-Pereira V., \& Opher R., 2000, ApJ, 534, 967

\bibitem{} Voit G. M., Donahue M., \& Slavin J. D., 1994, ApJ, 95, 87

\bibitem{}  Wentzel D. G., 1974, Sol. Phys., 39, 129

\end{thebibliography}
\end{document}